\documentclass[%
 reprint,
 amsmath,
 amssymb,
 aps,
 graphicx,
prb,
showpacs,
footinbib,
]{revtex4-1}

\usepackage{color}
\usepackage{multirow}
\usepackage{graphicx}% Include figure files
\usepackage{dcolumn}% Align table columns on decimal point
\usepackage{bm}% bold math
\begin{document}

\title{Surface organization of homoepitaxial InP films grown by metalorganic vapor-phase epitaxy}

\author{A. \surname{Gocalinska}}
\thanks{Corresponding author}
\affiliation{Tyndall National Institute, ``Lee Maltings'', University College Cork, Cork, Ireland}
\author{M. \surname{Manganaro}}
\affiliation{Tyndall National Institute, ``Lee Maltings'', University College Cork, Cork, Ireland}
\author{D. D. \surname{Vvedensky}}
\affiliation{The Blackett Laboratory, Imperial College London, London SW7 2AZ, United Kingdom}
\author{E. \surname{Pelucchi}}
\affiliation{Tyndall National Institute, ``Lee Maltings'', University College Cork, Cork, Ireland}

\date{\today}

\begin{abstract}
We present a systematic study of the morphology of homoepitaxial InP films grown by metalorganic vapor-phase epitaxy which are imaged with {\it ex situ} atomic force microscopy. These films show a dramatic range of different surface morphologies as a function of the growth conditions and substrate (growth temperature, V/III ratio, and miscut angle $<0.6^\circ$ and orientation toward A or B sites), ranging from stable step flow to previously unreported strong step bunching, over 10~nm in height. These observations suggest a window of growth parameters for optimal quality epitaxial layers. We also present a theoretical model for these growth modes that takes account of deposition, diffusion, and dissociation of molecular precursors, and the diffusion and step incorporation of atoms released by the precursors.  The experimental conditions for step flow and step bunching are reproduced by this model, with the step bunching instability caused by the difference in molecular dissociation from above and below step edges, as was discussed previously for  GaAs (001).

\end{abstract}


\maketitle

\section{Introduction}
\label{sec1}

Homoepitaxial films of InP are commonly grown as buffer and cladding/waveguiding layers, setting the base for numerous types of semiconductor devices.\cite{wada99,zhang09}  The morphology of these films can have a crucial impact on the quality of the overgrown material. For example, poor quality buffers could result in structural defects propagating to the device layers, reducing the carrier mobility or affecting optical properties, and thereby degrading the performance of the device.\cite{achard99,andreani03}

In general, an epitaxial process is required to fabricate materials with reproducible properties of the highest quality in terms of crystal structure, purity, alloy composition, surface morphology, etc. In most cases, material quality is controlled by the adjustment of the growth conditions, such as substrate temperature, gas/molecular flow and their ratios which, in metalorganic vapor-phase epitaxy (MOVPE), will affect precursors diffusion, surface decomposition processes, nucleation, diffusion of adatoms and their insertion at specific crystallographic sites and steps.\cite{carra98}  For molecular-beam epitaxy, a well-founded conceptual and computational framework has emerged \cite{evans06} since the pioneering work of Burton, Cabrera, and Frank (BCF).\cite{burton51}  Systematic studies based on surface diffraction measurements and scanning tunneling microscopy, in conjunction with extensive computational modeling have identified and characterized many of the atomistic processes, even in the complex setting of III-V systems.\cite{shitara92,avery97,itoh98,kratzer99,morgan99,kratzer02} The corresponding development for MOVPE has been comparatively slow, due in part to the limited availability of {\it in situ} measurements.\cite{richter02}  Nevertheless, systematic {\it ex situ} measurements of growth morphologies, again in conjunction with theoretical modeling, have revealed the importance of precursor diffusion and decomposition at step edges in determining the nature and scale of surface morphologies.\cite{chua08}

Epitaxy-ready substrates are prepared by cutting individual slices out of the bulk crystal at the desired angle with respect to the main crystallographic plane.\cite{mcp}  In the case of InP, the crystallographic steps exposed after this cutting may be ideally terminated either with indium (so called A-steps, with the substrate misoriented toward the [111]A planes), or phosphorus atoms (B-steps, with the substrate misoriented toward the [111]B  planes), or any combination of these two.\cite{doscher11,raghavachari02,fu02} Depending on the miscut, the exposed surface can be close to singular or contain a dense periodic array of crystallographic steps. Epitaxial overgrowth on such substrates may then proceed (when truly ``epitaxial'') in one of three basic growth modes:~step flow, where the overgrowing layer is advancing each exposed step at the same rate, creating an exact copy of the underlying surface; step bunching, where terrace formation is observed due to clustering of individual steps; or island formation, when the growth of new layer is initiated only at the step edges, but also between steps, in which case no long range order is observed. If the growth conditions are not optimal, due to defect formation, the surface morphology can be corrupted and will not resemble any of the three main modes. For most applications, step flow is the desired mode, as it leads to creation of atomically flat layers, without irregularities (such as, in an extreme example, antiphase boundaries).\cite{bronikowski93}

On the other hand, the role of small substrate misorientations for the III-V system  have been recently revealed for arsenide alloys (e.g. GaAs, AlInAs or InAs) grown by MOVPE,\cite{moret06,pelucchi06,chua08,young10,gocalinska12} where complex surface organization, striking effects and state-of-the-art results on the material optical/transport properties have been reported. In those reports, small changes of growth conditions significantly affected the epitaxial morphology. As usual, the growth conditions revolved around a few parameters, where the common variables were substrate miscut, growth temperature, growth rate and molar ratio between precursors injected into the reactor chamber (reactor pressure, carrier gas and precursor choice were fixed in the given setup).  Crucially, the authors found a strong correlation between surface organization and material properties.  What is even more relevant is that the authors linked the unusual variety of surface organization (e.g.~from islanding, through step flow to periodic step bunches just by varying substrate miscut), to the intrinsic two-step MOVPE growth process: first adsorption, diffusion and decomposition of molecular precursors at step edges and, subsequently, adatom diffusion and incorporation. Their model reproduced the condition for island nucleation as well as for the step bunching instabilities, linking this last process to the difference in molecular dissociation from above and below step edges.\cite{chua08}

Despite the broad and primary commercial interest and exploitation of InP,\cite{ogawa98} only limited data are available in the literature on InP grown by MOVPE. Only a small range of parameters/surface organization appear to have been correlated. For instance, the influence of small differences of substrate misorientation was partly discussed in Ref.~[\onlinecite{huang11}] (for substrates with offcuts of 0.02$^\circ$--0.25$^\circ$ toward (01$\bar{1}$)) and Ref.~[\onlinecite{cederberg11}] (for singular wafers and with misorientations of 0.2$^\circ$ toward [110], [111]A and [111]B). Morphologies created during homoepitaxy on singular wafers (with misorientation below 0.1$^\circ$) were also investigated in Ref.~\onlinecite{epler11} with respect to the direction of the miscut. In Ref.~[\onlinecite{epler11}], the authors analyzed the effect of temperature on step bunching on vicinal surfaces in a larger miscut range (from 0.2$^\circ$ to 2$^\circ$ toward [111]A). Nevertheless, there is no comprehensive study of the organization of the MOVPE-induced crystallographic steps on vicinal surfaces, so a fundamental understanding of the fundamental kinetic processes is still lacking.

In this paper, we report a systematic study of the impact of growth conditions on the surface morphology of InP surfaces with small misorientations during MOVPE. We show several surprising effects and a remarkably broad variety of surface organizational modes, well beyond what has been observed in the arsenide (GaAs) counterpart.  The organization of our paper is as follows.  The procedure used for our experiments is described in Sec.~\ref{sec2}.  The surface morphologies observed on vicinal InP(001) are described in Sec.~\ref{sec3}, with the presentation of the results grouped according to growth rate and V/III ratio (Secs.~\ref{sec3A}, \ref{sec3B}, \ref{sec3C}) and the effect of doping (Sec.~\ref{sec3D}).  These results are analyzed in Sec.~\ref{sec4}, which includes a theoretical discussion based on an extension of the model in Ref.~[\onlinecite{chua08}].  One of the interesting aspects of our experiments is that we can examine the effect of the group-V species on the various steps of the growth kinetics.  We summarize our results in Sec.~\ref{sec5}, where we also discuss future modelling strategies.

\section{Experimental procedure}
\label{sec2}

All samples in this study were grown by MOVPE at low pressure (80~mbar) in a commercial horizontal reactor with purified N$_2$ as the carrier gas and trimethylindium (TMIn) and phosphine (PH$_3$) as precursors.\cite{dimas11}  300-nm-thick single layers of InP were grown on perfectly oriented or slightly misoriented InP(001) substrates, which in most cases were semi-insulating Fe-doped, with reference samples grown on Zn- ($p$-type) and Si-doped ($n$-type) wafers (we anticipate no observed differences to result from the substrate doping, as no diffusion to the epitaxial layer is expected). The majority of the InP layers were grown as nominally undoped. The unintentional doping level is estimated to be well below $5\times10^{15}~\mbox{cm}^{-3}$, which is our detection threshold.  We have also grown several samples using diethylzinc (DEZn) and disilane (Si$_2$H$_6$) as precursors to investigate the impact they have on the surface morphology when incorporated into the InP matrix. Doping levels were $5\times10^{17}~\mbox{cm}^{-3}$ for Zn and $9\times10^{17}~\mbox{cm}^{-3}$ for Si and constant across the layer, as confirmed by electrochemical capacitance voltage measurements. Such carrier concentrations are considered moderate for InP-based semiconductor devices.\cite{choi09}

Growth conditions were varied systematically across samples:~the estimated actual growth temperature $T_G$ varied in the range 520$^\circ$C to 720$^\circ$C, the V/III ratio $R$  from 150 to 450, and sample misorientation from nominally ``perfectly oriented'' to 0.6$^\circ$ (all with a tolerance of $\pm0.02^\circ$)  toward [111] A or B. The growth rate $G$ was kept constant at 0.7~$\mu$m/hr for most of the samples, with reference/comparison growth carried out at 0.35~$\mu$m/hr and 1.4~$\mu$m/hr. All together, more than 150 samples were grown in various conditions for this study, with several control growth runs performed to ensure the full reproducibility of the reported results. In all cases, particular attention was paid to reactor environment quality and to temperature control by growing the samples only with reactor walls already baked and covered by previous growth runs. The growth temperature was estimated by emissivity corrected pyrometry. The relevant details are referenced below when a given example is discussed.

\begin{table*}[t!]
\renewcommand{\arraystretch}{1.1}
\caption{\label{table1}The growth rate, V/III ratios, misorientation, and growth temperature of InP substrates produced during homoepitaxy by MOVPE.  The samples are labeled for each set of growth conditions, together with an abbreviation for the observed morphology: discrete islands (I), diffuse islands (DI), step flow (SF), ``normal'' step bunching (SB), cliffs (C), braids (B), and defects (D).}\medskip
\begin{tabular}{|c|c|c|c|c|c|c|c|c|c|c|}
\hline
\multirow{2}{*}{Growth rate} & \multirow{2}{*}{V/III ratio} & \multirow{2}{*}{Substrate} & \multicolumn{8}{c|}{Growth temperature} \\
\cline{4-11}
& & & 520$^\circ$C & 560$^\circ$C & 585$^\circ$C & 610$^\circ$C & 630$^\circ$C & 655$^\circ$C & 685$^\circ$C & 720$^\circ$C \\
\hline
\multirow{20}{*}{0.7~$\mu$m/hr} & \multirow{10}{*}{150} & \multirow{2}{*}{``perfectly oriented''} &  {\bf 1A} &  {\bf 2A} &  {\bf 3A} &  {\bf 4A} & {\bf 5A} & {\bf 6A} & {\bf 7A} &  {\bf 8A}  \\
& & & I & I & I & SF & SB & SB & SB &  D \\
\cline{3-11}
& & \multirow{2}{*}{0.2$^\circ$ toward [111]A} & {\bf 1B} & {\bf 2B} & {\bf 3B} & {\bf 4B} &  {\bf 5B} &  {\bf 6B} & {\bf 7B} & {\bf 8B}  \\
& & & DI &  DI & C & C & C & B & D & D \\
\cline{3-11}
& & \multirow{2}{*}{0.4$^\circ$ toward [111]A} & {\bf 1C} & {\bf 2C} & {\bf 3C} & {\bf 4C} & {\bf 5C} &  {\bf 6C} & {\bf 7C} &  {\bf 8C}  \\
& & & DI & DI & C & C & C & B & D & D \\
\cline{3-11}
& & \multirow{2}{*}{0.4$^\circ$ toward [111]B} & {\bf 1D} & {\bf 2D} & {\bf 3D} & {\bf 4D} & {\bf 5D} & {\bf 6D} & {\bf 7D} & {\bf 8D}  \\
& & & DI & DI & C & C & B & D & D & D \\
\cline{3-11}
& & \multirow{2}{*}{0.6$^\circ$ toward [111]A} & {\bf 1E} & {\bf 2E} & {\bf 3E} & {\bf 4E} & {\bf 5E} & {\bf 6E} & {\bf 7E} & {\bf 8E}  \\
& & &  DI & DI & C & C & B & SB & D & D \\
\cline{2-11}
& \multirow{10}{*}{450} & \multirow{2}{*}{``perfectly oriented''} &  {\bf 9A} & \multirow{2}{*}{---} & {\bf 10A} & \multirow{2}{*}{---} & {\bf 11A} & {\bf 12A} & \multirow{2}{*}{---} & {\bf 13A}   \\
& & & SF & & SF & & SB & SB & & SF \\
\cline{3-11}
& & \multirow{2}{*}{0.2$^\circ$ toward [111]A} &  {\bf 9B} & \multirow{2}{*}{---} & {\bf 10B} & \multirow{2}{*}{---} & {\bf 11B} & {\bf 12B} & \multirow{2}{*}{---} & {\bf 13B}   \\
& & & SB & & B & & B & B & & SB \\
\cline{3-11}
& & \multirow{2}{*}{0.4$^\circ$ toward [111]A} &  {\bf 9C} & \multirow{2}{*}{---} & {\bf 10C} & \multirow{2}{*}{---} & {\bf 11C} & {\bf 12C} & \multirow{2}{*}{---} & {\bf 13C}   \\
& & & SB & & B & & B & B & & SB \\
\cline{3-11}
& & \multirow{2}{*}{0.4$^\circ$ toward [111]B} &  {\bf 9D} & \multirow{2}{*}{---} & {\bf 10D} & \multirow{2}{*}{---} & {\bf 11D} & {\bf 12D} & \multirow{2}{*}{---} & {\bf 13D}   \\
& & & SF & & SF & & SB & SB & & SF \\
\cline{3-11}
& & \multirow{2}{*}{0.6$^\circ$ toward [111]A} &  {\bf 9E} & \multirow{2}{*}{---} & {\bf 10E} & \multirow{2}{*}{---} & {\bf 11E} & {\bf 12E} & \multirow{2}{*}{---} & {\bf 13E}   \\
& & & SB & & B & & B & B & & SB \\
\hline
\end{tabular}
\end{table*}

\begin{table}[t!]
\renewcommand{\arraystretch}{1.1}
\caption{\label{table2}The growth rate, V/III ratios and misorientation of InP substrates for homoepitaxial growth at $T_G=630^\circ$C during MOVPE.  The samples are labeled for each set of growth conditions, together with an abbreviation for the observed morphology used in Table~\ref{table1}.}\medskip
\begin{tabular}{|c|c|c|c|}
\hline
\multirow{2}{*}{Substrate} & $G=0.35~\mu$m/hr & \multicolumn{2}{c|}{$G=1.4~\mu$m/hr} \\
\cline{2-4}
& $R=450$ & $R=150$ & $R=450$ \\
\hline
\multirow{2}{*}{``perfectly oriented''} & {\bf 14A} & {\bf 15A} & {\bf 16A} \\
& SF & SB & SF \\
\hline
\multirow{2}{*}{0.2$^\circ$ toward [111]A} & {\bf 14B} & {\bf 15B} & {\bf 16B} \\
& B & B & SB \\
\hline
\multirow{2}{*}{0.4$^\circ$ toward [111]A} & {\bf 14C} & {\bf 15C} & {\bf 16C} \\
& B & B & SB \\
\hline
\multirow{2}{*}{0.4$^\circ$ toward [111]B} & {\bf 14D} & {\bf 15D} & {\bf 16D} \\
& SF & B & SF \\
\hline
\multirow{2}{*}{0.6$^\circ$ toward [111]A} & {\bf 14E} & {\bf 15E} & {\bf 16E} \\
& SB & B & SB \\
\hline
\end{tabular}
\end{table}

\begin{table}[t!]
\renewcommand{\arraystretch}{1.1}
\caption{\label{table3}Summary of surface morphologies of the samples in Tables~\ref{table1} and \ref{table2}.}\medskip
\begin{tabular}{|c|c|lllll|}
\hline
\multicolumn{2}{|c|}{Surface morphology} & \multicolumn{5}{c|}{Sample number} \\
\hline
\multirow{3}{*}{Islands} & Discrete (I) & 1A, & 2A,  & 3A && \\
\cline{2-7}
& \multirow{2}{*}{Diffuse (DI)} & 1B, & 1C, & 1D, & 1E, & 2B, \\
& & 2C, & 2D, & 2E &&  \\
\hline
\multirow{6}{*}{``Plain''} & \multirow{3}{*}{Step flow (SF)} & 4A, & 9A, & 9D, & 10A, & 10D, \\
& & 13A, & 13D, & 14A, & 14D, & 16A, \\
& & 16D &&&& \\
\cline{2-7}
& \multirow{3}{*}{``Normal'' step} & 5A, & 6A, & 6E, & 7A, & 9B, \\
& \multirow{3}{*}{ bunching (SB)} & 9C, & 9E, & 11A, & 11D, & 12A,  \\
& & 12D, & 13B, & 13C, & 13E, & 14E,  \\
& & 15A, & 16B, & 16C, & 16C, &16E \\
\hline
\multirow{3}{*}{Step bunching} & \multirow{2}{*}{Cliffs (C)} & 3B, & 3C, & 3D, & 3E, & 4B, \\
\multirow{3}{*}{with} & & 4C, & 4D, & 4E, & 5B, & 5C \\
\cline{2-7}
\multirow{3}{*}{long-range} & \multirow{4}{*}{Braids (B)} & 5D, & 5E, & 6B, & 6C, & 10B, \\
\multirow{3}{*}{periodicity} & & 10C, & 10E, & 11B, & 11C, & 11E, \\
& & 12B, & 12C, & 12E, & 14B, & 14C, \\
& & 15B, & 15C, & 15D, & 15E & \\
\hline
\multirow{2}{*}{Defects} & \multirow{2}{*}{Defects (D)} & 6D, & 7B, & 7C, & 7D, & 7E, \\
& & 8A, & 8B, & 8C, & 8D, & 8E \\
\hline
\end{tabular}
\end{table}

All epitaxial growth resulted in mirror-like surfaces. The samples were first inspected with an optical microscope in (Nomarski) differential interference contrast (NDIC), as well as in dark-field mode. The vertical resolution of NDIC is limited and a subsequent detailed morphological study was performed with atomic force microscopy (AFM) in tapping/non contact mode at room temperature and in air. The results are presented in various scan sizes to emphasize the features relevant to a particular growth morphology. While we always measured our samples at a short interval from removal from the reactor, the surface features did not appear do degrade significantly over several weeks, an indication that, as for GaAs, the surface oxide develops slowly enough and conformally to the original structure in the first months of air exposure. The quality/crystallinity of the grown material was also confirmed by X-ray diffraction measurements.

\section{Results}
\label{sec3}

Our experiments revealed a surprising variety of InP surface morphological organization during homoepitaxy by MOVPE. The growth parameters, which are the growth rate $G$, V/III ratio $R$, misorientation angle and direction, and growth temperature $T_G$ for each sample are compiled in Tables \ref{table1} and \ref{table2}, with labels that will be used in the following discussion. The samples associated with each type of morphology are listed in Table \ref{table3}.  We first discuss the results for the growth of nominally undoped layers for different growth rates and V/III ratios, before considering the effect of doping.

\subsection{$\bm{G=0.7~\mu}$m/hr, $\bm{R=150}$}
\label{sec3A}

\begin{figure}[t!]
\includegraphics[width=8.5cm]{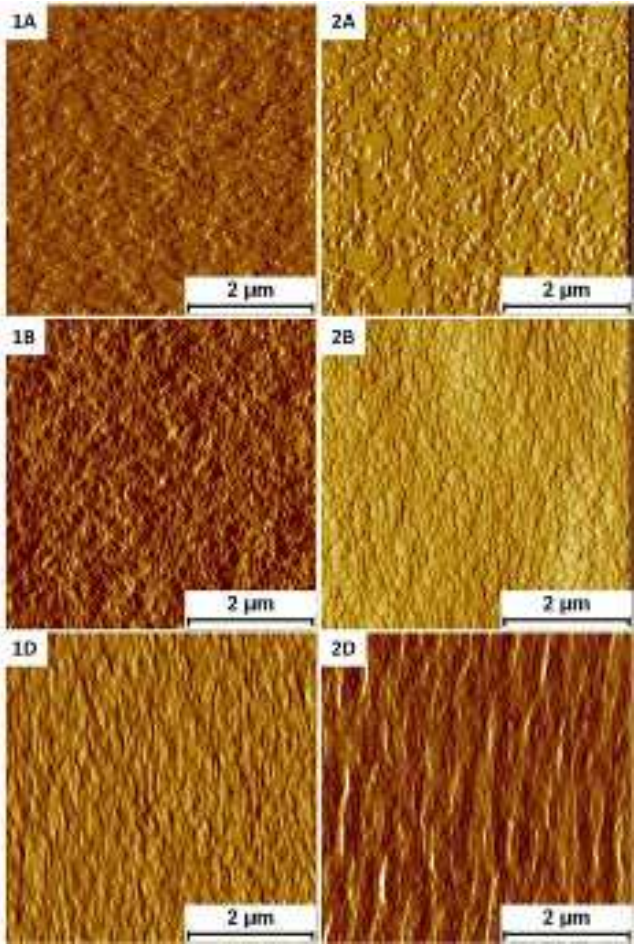}
\caption{\label{fig1}(Color online) AFM images (signal amplitudes) of the top surface of samples grown with $R =150$, at $T_G=520^\circ$C (left panel) and $T_G=560^\circ$C (right panel), on substrates that are perfectly oriented $(\pm0.02^\circ)$ (1A, 2A), misoriented by 0.2$^\circ$ toward [111]A (1B, 2B) and by 0.4$^\circ$ toward [111]B (1D, 2D).}
\end{figure}

\begin{figure}
\includegraphics[width=8.7cm]{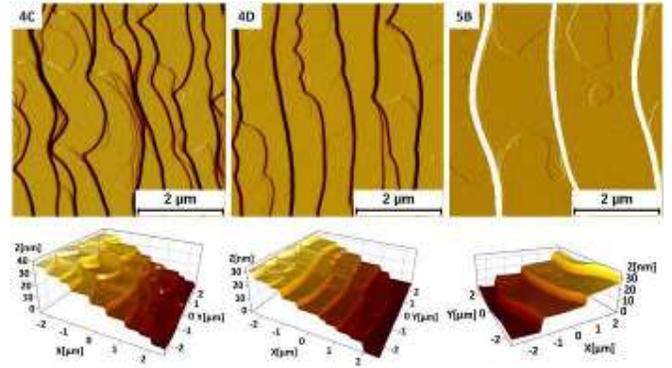}
\caption{\label{fig2}(Color online) AFM images (signal amplitudes in top row and corresponding three-dimensional (3D) height reconstructions in bottom row) of samples showing cliff-like step-bunching. Growth was performed on substrates with a misorientation of 0.4$^\circ$ toward [111]A (4c), 0.4$^\circ$ toward [111]B (4D), and 0.2$^\circ$ toward [111]A  (5B) at growth conditions with $R=150$, $G= 0.7~\mu$m/hr, $T_G=610^\circ$C (4C, 4D) and $T_G=630^\circ$C (5B).}
\end{figure}

On samples 1A-3A (low $T_G$, no intentional misorientation), we observed the random nucleation of distinct and separated epitaxial islands  between the crystallographic steps (when discernible) of the substrate (Fig.~\ref{fig1}). This morphology will be referred to as ``islanding'' and signified by I. On samples 1B-1E and 2B-2E  (low $T_G$, with intentional substrate misorientation), however, a disorganized mixture of step flow and step bunching developed on which the steps of the original substrate could not be easily discerned  (Fig.~\ref{fig1}).  This morphology, called ``diffuse islanding'', will be signified by DI.  This type of morphology is similar to that observed in models and scans with scanning tunnelling of growth on misoriented surfaces during MBE\cite{shitara92,wolf97} in a transition regime between island nucleation and growth and step flow.  When the growth temperature was raised we observed more organized surface features. The samples grown on perfectly oriented wafers showed morphologies close to step flow, but with some residual islands (4A), step bunching of several neighboring steps without significant long range organization (5A), or in close grouping of two subsequent steps (6A.)  The step-bunched morphology will be denoted as SB.

On samples with an intentional misorientation, significant step bunching was observed at lower temperatures than for the perfectly oriented surfaces. Samples 3B-3E, 4B-4E and 5B-5C showed the formation of ``cliffs'', denoted as C, made of up to steep 40 monolayer stacks (corresponding to 10-nm-high edges). The general trend is similar for all these samples, but the observed features were the most distinct on samples 3D, 4B-4E and 5B (Fig.~\ref{fig2}).  When investigated in detail, the cliff morphology shows a very dense succession of nm step bunches, rather than atomically sharp edges.

\begin{figure}
\includegraphics[width=8.7cm]{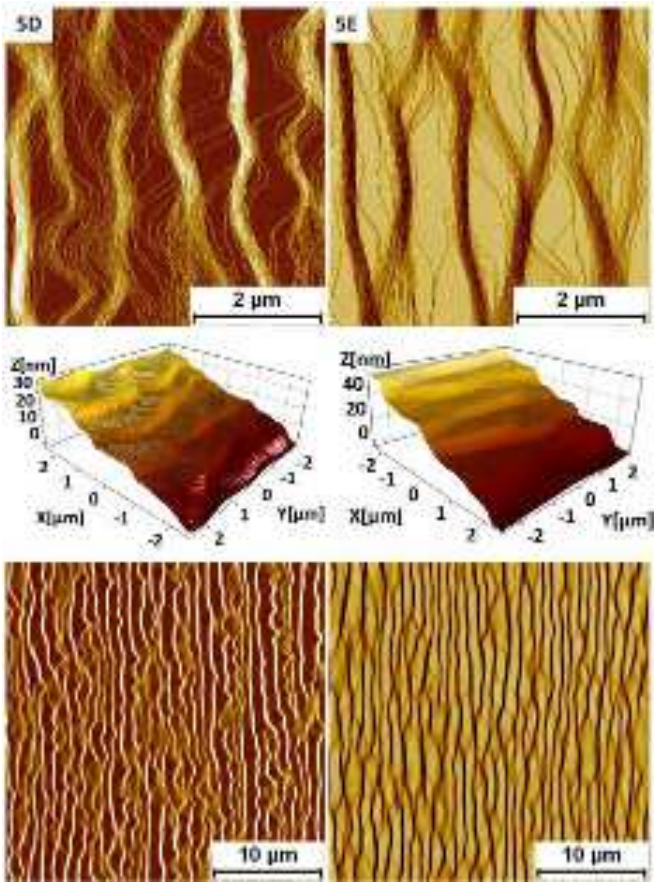}
\caption{\label{fig3}(Color online) AFM images (signal amplitudes in top and bottom row and 3D height reconstruction in central row) of samples showing braid-like step bunching. The samples were grown on substrates with a misorientation of 0.4$^\circ$ (5D) and 0.6$^\circ$ toward [111]A (5E) with $R=150$ and $G= 0.7~\mu$m/hr at $T_G=630^\circ$C. The top and central rows contain images in close zoom, while the bottom row shows the large-scale organization of the features.}
\end{figure}

Different morphologies were obtained for $T_G> 630^\circ$C. This was first evident for samples with largest misorientation (5D, 5E), and then at $T_G=655^\circ$C for all A-misoriented samples (6B, 6C, 6E), while B-misoriented (6D) and perfectly-oriented (6A) samples showed moderate step bunching, but with an indication of defect formation. Strongly step-bunched surfaces were still observed, but the bunched steps were not simply overlapping;~instead, while maintaining a closely grouped and periodic structure, they showed significant numbers of single steps between the bunches, creating a more ``braid-like'' picture on a flattened image  (Fig.~\ref{fig3}). Moreover, on a scale of 10--30 $\mu$m,  the ``braids''  were seen to dissolve into single steps which subsequently plaited into the next ``braid'' ahead.  This morphology will be signified by B.

\begin{figure}
\includegraphics[width=8.7cm]{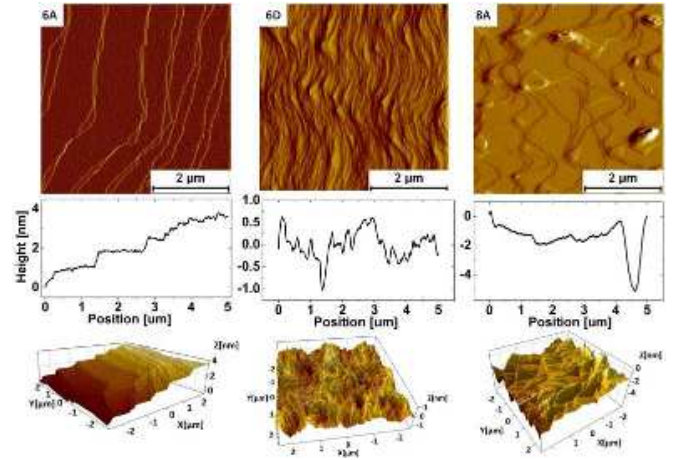}
\caption{\label{fig4}(Color online) AFM images (signal amplitudes in top row, corresponding height profiles in central row and 3D height reconstructions in bottom row) of sample surfaces. Images refer to samples grown at $R=150$, $G=0.7~\mu$m/hr, $T_G=655^\circ$C (6A, 6D) and $T_G=720^\circ$C (8A), using substrates that are perfectly-oriented (6A, 8A) or misoriented by 0.4$^\circ$ toward [111]B (6D). Samples 6D and 8A show defected sample surfaces, while the sample 6A shows step pinning (sharp change in the step edge line).}
\end{figure}

For the highest growth temperatures, we observed significant disruption of the sample surfaces. Indentations with a depth of up to 5~nm, corresponding to approximately 20~monolayers (MLs), were measured on samples 8A-8E, with smaller dimples and step pinning observed also at slightly lower temperatures (sample 6A, 6D and 7A-7E). Plan view images, lines scans, and three-dimensional images of such surfaces are shown in Fig.~\ref{fig4}.  These defected surfaces are signified by D.

\begin{figure}
\includegraphics[width=8.7cm]{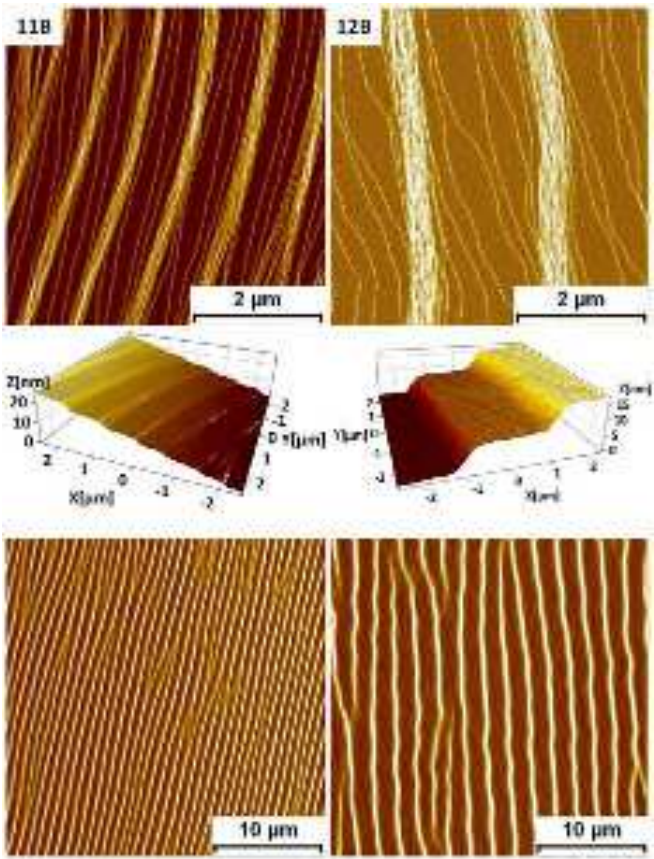}
\caption{\label{fig5}(Color online) AFM images (signal amplitudes in top and bottom row and 3D height reconstruction in central row) of samples showing braid-like step bunching. These images refer to samples grown on substrates with an off-cut of 0.2$^\circ$ toward [111]A at $R=450$, $G= 0.7~\mu$m/hr, $T_G=630^\circ$C (11B) and $T_G=655^\circ$C (12B). The top and central rows contain images in close zoom, while the bottom row shows the large-scale organization of the features.}
\end{figure}

\begin{figure}
\includegraphics[width=8.7cm]{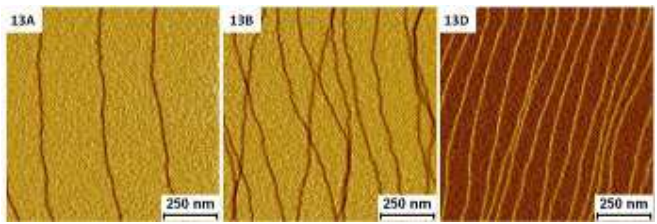}
\caption{\label{fig6}(Color online) AFM images (signal amplitudes) of the top surface of samples grown at 720$^\circ$C and $R=450$ on substrates that are perfectly oriented (13A) and misoriented by 0.2$^\circ$ toward [111]A (13B) and 0.4$^\circ$ toward [111]B (13D).}
\end{figure}

\subsection{$\bm{G=0.7~\mu}$m/hr, $\bm{R=450}$}
\label{sec3B}

Samples grown under higher phosphine flow showed a much less diverse range of morphologies than those with lower flows. Starting from growth temperatures as low as 520$^\circ$C we were able to identify organized step behavior on the surfaces of all samples. The singular and B-misoriented samples (9A, 9D) showed step-flow, while the A-misoriented samples showed normal ``disordered''  step bunching (with step jumps of a few MLs, up to 2~nm) without long-range periodicity (9B-9C, 9E). Nevertheless, a slight increase in the growth temperature resulted in long-range step organization. While singular and B-misoriented samples (10A-12A and 10D-12D) seemed not to be affected, all of the A-misoriented samples (10B-10C, 10E) showed significant step bunching, which became even more pronounced at higher temperatures (11B-11C, 11E, 12B-12C, 12E), resembling the morphologies in Figs.~\ref{fig2} and \ref{fig3}. We again observed the plaiting braid-like morphology, even with increased ordering, which remained up to our highest growth temperature ($T_G=720^\circ$C), as shown in Fig.~\ref{fig5}.

Contrary to what was previously observed (defected areas), in these growth conditions, the high growth temperature resulted in very smooth surfaces, with step flow or limited/normal step-bunching on high off-cut wafers (Fig.~\ref{fig6}), which might be an indication that the previously observed indentation might be related to an insufficient supply of phosphorus to the surface.

\subsection{$\bm{G=1.4/0.35~\mu}$m/hr, $\bm{R=450/150}$}
\label{sec3C}

\begin{figure}[t!]
\includegraphics[width=8.7cm]{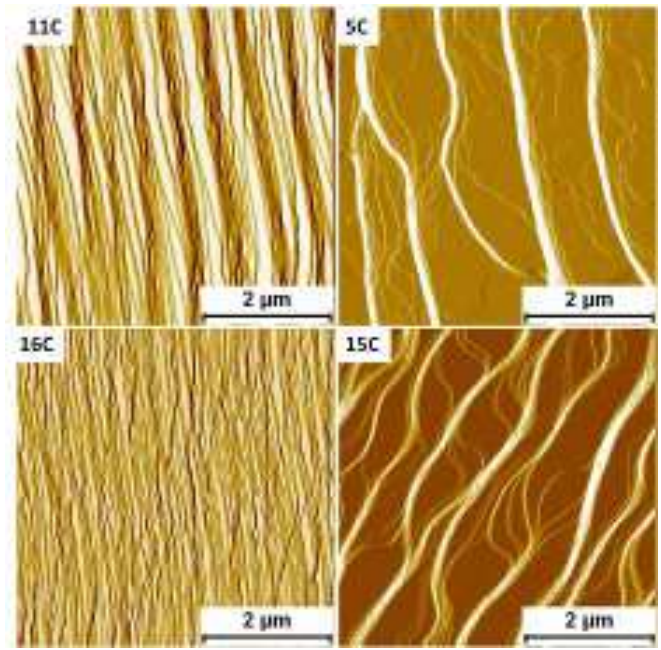}
\caption{\label{fig7}(Color online) AFM images (signal amplitudes) of the top surface grown on samples misoriented by 0.4$^\circ$ toward [111]A at $R=450$ (11C) and $R=150$ (5C) at $G=0.7\mu$m/hr, and $G=1.4\mu$m/hr (16C, 15C). Bunched areas in samples 5C and 15C are $\sim$20~MLs in height with a shape change. 5C shows more sharp cliffs, while 15C is more slope-like.}
\end{figure}

The effect of varying the growth rate and the phosphine flow rate was investigated only at $T_G=630^\circ$C, with the morphologies summarized in Table~\ref{table2}. Notably, when the growth rate was increased, we observed an improvement in the uniformity of the surface. The effects of the change (at $G = 1.4~\mu$m/hr compared to our standard of $G=0.7~\mu$m/hr) was consistent for samples with both high ($R=450$) and low ($R=150$) phosphine flow rates:~the large step bunching was softened, resulting in the absence of long-range periodicity for the samples grown with $R=450$, while a softening of the cliff-like edges on samples growth with $R=150$ (Fig.~\ref{fig7}). On the other hand, reducing the growth rate to 0.35~$\mu$m/hr seemed to have no effect on the morphology compared to samples grown with $0.7~\mu$m/hr.

\begin{figure}
\includegraphics[width=8.7cm]{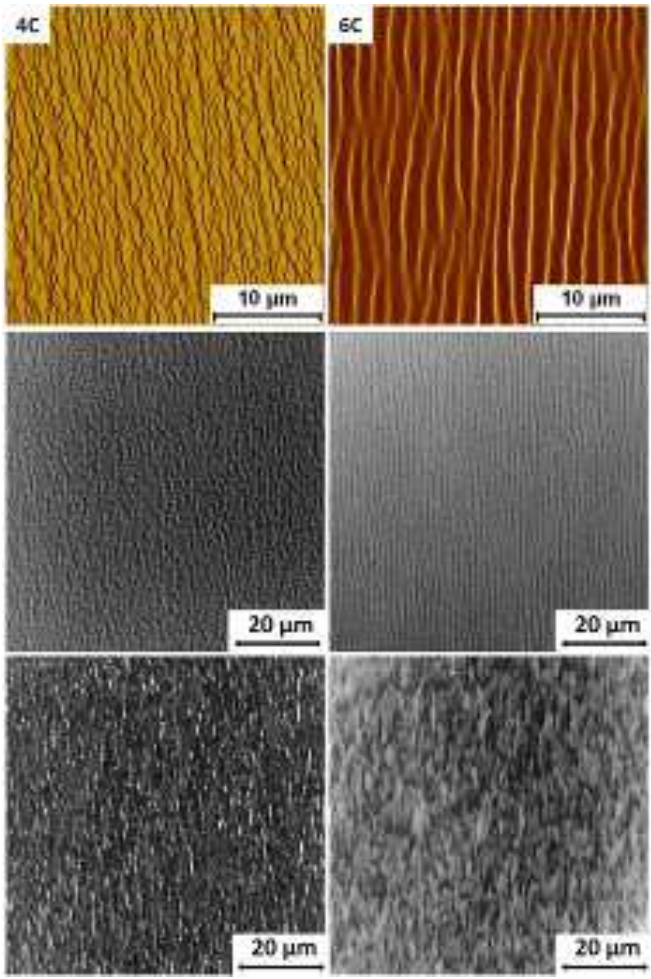}
\caption{\label{fig8}(Color online) Long-range organization of the surfaces of samples 4C and 6C imaged by AFM (signal amplitude, top row), N-DIC microscopy (middle row) and dark-field optical microscopy (bottom row). Optical and AFM images are of different scale, as indicated by the scale bars.}
\end{figure}

It is interesting to note that, as expected, the samples showing island formation, step flow and ``normal'' step bunching appear to be featureless under the optical microscope. Nevertheless, the significantly step-bunched samples had surface morphologies which enabled characterization with either dark-field microscopy or N-DIC. While the N-DIC images clearly showed the lines corresponding to the feature edges, the dark-field configuration accentuated the intertwining of the ``braid-like'' bunches (Fig.~\ref{fig8}). This is significant for standard laboratory practice, as often first quality control of the growth is performed under optical microscope. Thus it is important to note, that observed features do not necessarily correspond to defected layer and might be originating from surface morphology.

\subsection{Doping}
\label{sec3D}

\begin{figure}[t!]
\includegraphics[width=8.7cm]{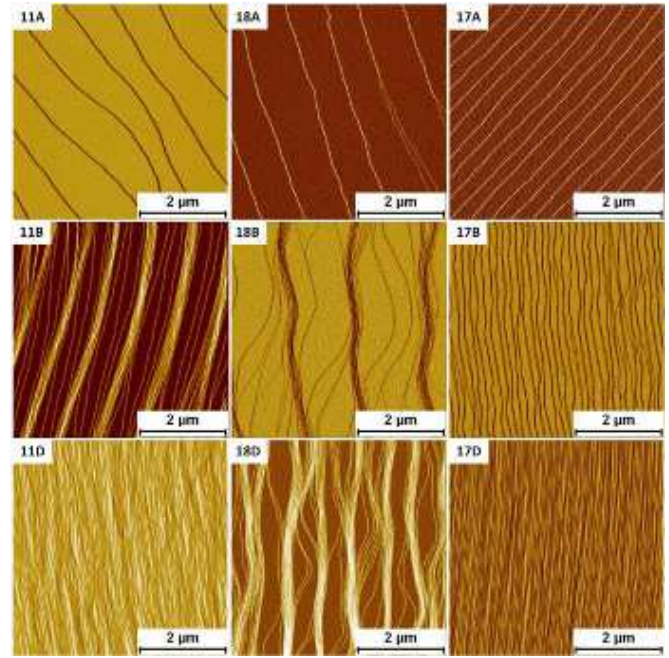}
\caption{\label{fig9}(Color online) AFM images (signal amplitudes) of the top surface of undoped (left panel), Si- and Zn-doped samples (central and right panel, respectively) grown on perfectly oriented substrates (top row), and substrates misoriented by 0.2$^\circ$ toward [111]A (middle row) and by 0.4$^\circ$ toward [111]B (bottom row).}
\end{figure}

Depending on the application, the structural design of a device often calls for conductive InP layers. This is realized by the intentional doping of the bulk materials, often with elements such as zinc (Zn) or silicon (Si). In InP the Si could in principle introduce either a donor state by replacing In, or an acceptor state by replacing P, but the donor state is more energetically favorable, and therefore dominant.\cite{sze85} The incorporation of Si is proportional to the disilane flow and the incorporation efficiency is high, i.e.~the silane flow needed to achieve our target doping levels is an order of magnitude lower than the TMIn flow.\cite{blaauw91} However, Zn doping normally has a very low incorporation efficiency,\cite{logan96,chu96} so less than 0.2\% of the molecules injected into the reactor were incorporated into the crystal matrix with a molar ratio TMIn/DeZn $\approx$ 50.

In Table~\ref{table4} we list the growth parameters for the samples discussed in this section. Figure~\ref{fig9} presents a comparison of the morphology of layers grown on semi-insulating substrates with those discussed in preceding sections that have the same misorientation.  For this study we used $R=520$ (slightly different from previously used values, as optimized for different experiments in our laboratory, nevertheless comparable to the high V/III ratios reported before), $T_G=630^\circ$C and $G=0.7~\mu$m/hr. Good surface quality was obtained for all growth runs, as expected. However, differences were observed when step-step interactions are of interest.

\begin{table}[t!]
\renewcommand{\arraystretch}{1.1}
\caption{\label{table4}Misorientation and layer dopant for InP substrates at $T_G=630^\circ$C and $R=520$.  The samples are labeled for each set of growth conditions, together with an abbreviation for the observed morphology used in Table~\ref{table1}.}\medskip
\begin{tabular}{|c|p{1cm}|p{1cm}|}
\hline
\multirow{2}{*}{Substrate} & \multicolumn{2}{c|}{Dopant} \\
\cline{2-3}
& \multicolumn{1}{c|}{Zn} & \multicolumn{1}{c|}{Si} \\
\hline
\multirow{2}{*}{``perfectly oriented''} & \multicolumn{1}{c|}{\bf 17A} & \multicolumn{1}{c|}{\bf 18A} \\
& \multicolumn{1}{c|}{SF} & \multicolumn{1}{c|}{SB} \\
\hline
\multirow{2}{*}{0.2$^\circ$ toward [111]A} & \multicolumn{1}{c|}{\bf 17B} & \multicolumn{1}{c|}{\bf 18B} \\
& \multicolumn{1}{c|}{SF} & \multicolumn{1}{c|}{B} \\
\hline
\multirow{2}{*}{0.4$^\circ$ toward [111]A} & \multicolumn{1}{c|}{\bf 17C} & \multicolumn{1}{c|}{\bf 18C} \\
& \multicolumn{1}{c|}{SB} & \multicolumn{1}{c|}{B} \\
\hline
\multirow{2}{*}{0.4$^\circ$ toward [111]B} & \multicolumn{1}{c|}{\bf 17D} & \multicolumn{1}{c|}{\bf 18D} \\
& \multicolumn{1}{c|}{SB} & \multicolumn{1}{c|}{B} \\
\hline
\multirow{2}{*}{0.6$^\circ$ toward [111]A} & \multicolumn{1}{c|}{\bf 17E} & \multicolumn{1}{c|}{\bf 18E} \\
& \multicolumn{1}{c|}{SB} & \multicolumn{1}{c|}{B} \\
\hline
\end{tabular}
\end{table}

Remarkably, Zn-doped layers showed significantly less step-bunching, while the Si doping seemed to have less of an impact, even if it appeared to have intensified kink formation and affected growth on B-misoriented substrates more than that on A-misoriented substrates. 
While for undoped epilayers with similar growth conditions ($R=450$) we observed bunching of two steps on singular wafers, braid-like step bunching on A-type substrates, and normal step bunching on B-type substrates, the incorporation of Zn atoms led to ideal single step-flow on perfectly oriented substrates and normal/modest step bunching (of only a few monolayers) on all misoriented wafers, without any unusually large terrace formation. On the other hand, Si-doped layers showed ``braid''-like step bunching, even on B-type substrates, rarely seen in undoped layers (with the exception of sample 5D). The use of Zn- or Si-doped substrates (as we anticipated earlier in our contribution), seemed to have no effect on the surface morphology -- the features were consistent with those seen on semi-insulating substrates (not shown).

\section{Discussion}
\label{sec4}

\subsection{Terrace lengths and step bunches}
\label{sec4A}

\begin{figure}[t!]
\includegraphics[width=8.7cm]{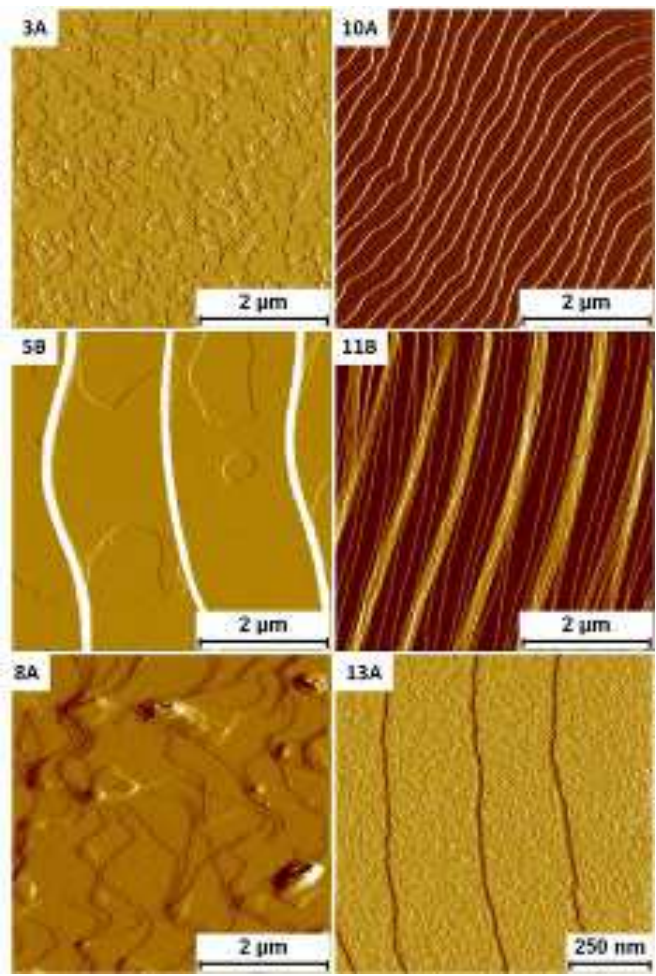}
\caption{\label{fig10}(Color online) Comparison of different morphologies as described in the main text.  AFM images (signal amplitudes) of the top surface of samples grown with $R=150$ (left panel) and $R=450$ (right panel), at $T_G=585^\circ$C (top row), 630$^\circ$C (middle row) and 720$^\circ$C (bottom row).}
\end{figure}

\begin{figure}[t!]
\includegraphics[width=8cm]{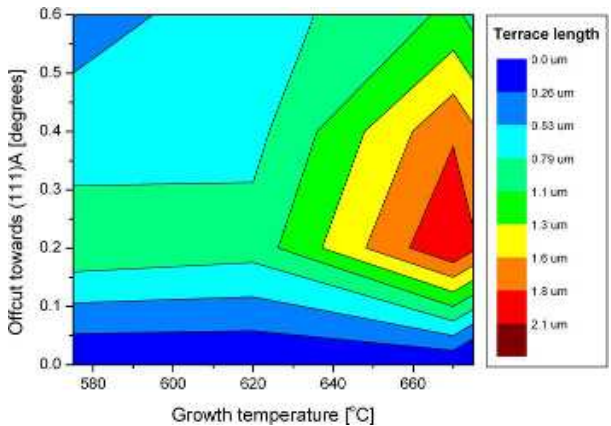}
\caption{\label{fig11}(Color online) Dependence of terrace lengths after growth on the initial substrate misorientation angle and growth temperature for samples grown with $R=450$ and $G=0.7~\mu$m/hr.}
\end{figure}

\begin{figure}[b!]
\includegraphics[width=8cm]{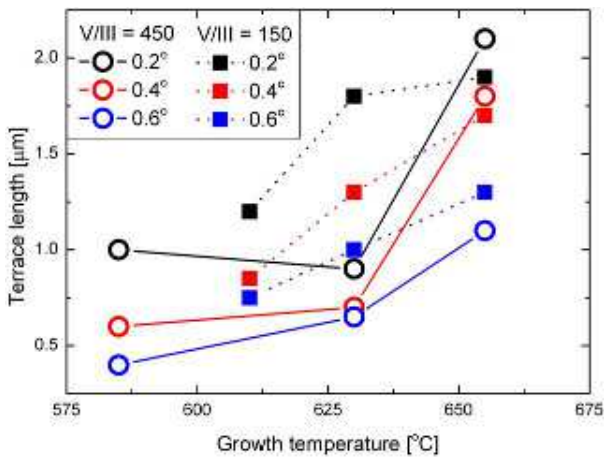}
\caption{\label{fig12}(Color online) Comparison of measured distances between bunches on samples grown with $R=450$ (circles) and $R=150$ (filled squares) on substrates with the indicated misorientation angles.}
\end{figure}

The wide range of growth conditions used in our investigation has revealed several identifiable trends in the resulting surface morphologies. The V/III ratio had the strongest impact on the morphology:~for growth at $R=150$, the temperature window for obtaining good surface morphology was extremely narrow. Growth temperatures lower than 585$^\circ$C were not sufficient to allow the formation of crystallographic steps while, at 655$^\circ$C, we already observed the appearance of surface defects (Fig.~\ref{fig10}). Increasing to $R=450$ expanded the allowed range to essentially all temperatures we used -- no defects were observed even at 720$^\circ$C, nor were the surfaces produced at 520$^\circ$C significantly worse than those at higher temperatures (Fig.~\ref{fig10}).  We observed a lengthening of the terraces with temperature and, of course,  a decrease of this length with larger substrate misorientation angle. (Figs.~\ref{fig11} and \ref{fig12}).  An increase of the phosphine flow to $R=650$ did not bring about any observable change in surface morphology.   We also observed good results for growth with $R=370$ (not shown). Cliff-like step bunching was observed only for lower phosphine flows.

\begin{figure*}[t!]
\includegraphics[width=16cm]{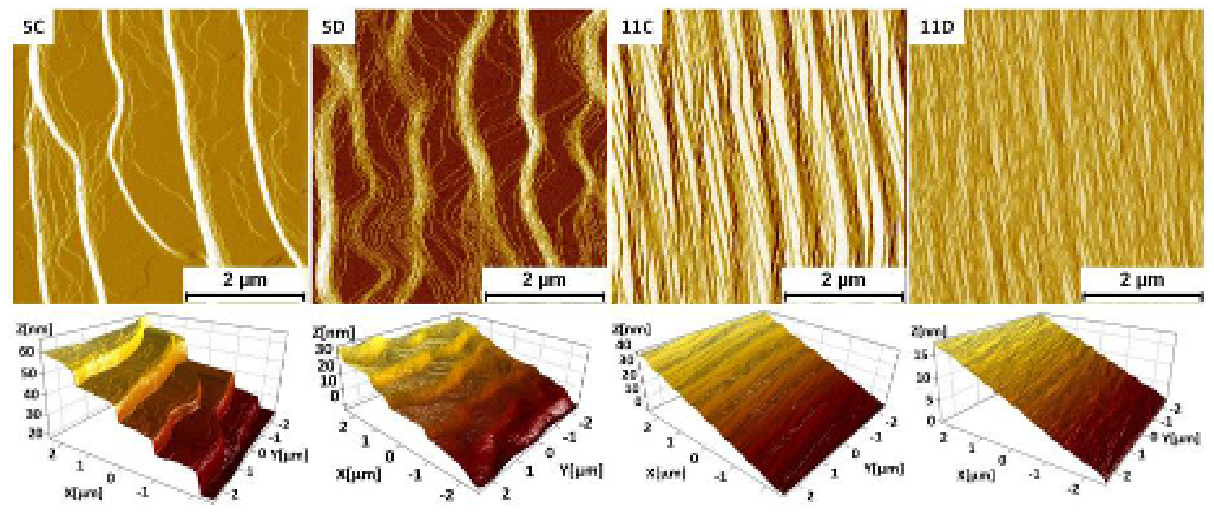}
\caption{\label{fig13}(Color online) AFM images (signal amplitudes in the top row, 3D height reconstructions in bottom row) of samples grown with $G = 0.7\mu$m/hr at $T_G=630^\circ$C and $R=150$ (5C, 5D) or $R=450$ (11C, 11D) on substrates misoriented by 0.4$^\circ$ toward [111]A (5C, 11C) or [111]B (5D, 11D).}
\end{figure*}

The growth rate also has a significant impact on the surface features.  The terrace formation observed on layers grown at $G=0.35~\mu$/hr and $0.7~\mu$m/hr was transformed into more conventional step bunching or step flow for $G=1.4\mu$m/hr and $R=450$, and the cliff-like kinks were ``softened'' into braids for $R=150$ (Fig.~\ref{fig7}). We nevertheless caution the reader that, as we have already mentioned, at high growth rates, we limited our studies to a specific growth temperature, and that further work is necessary to draw broader conclusions.

The defects observed for growth at high temperatures could be easily associated with desorption of phosphine from the surface. What supports that conclusion is the fact that higher phosphine flow cured the sample surface, creating good morphologies even at very high growth temperatures. Additionally, the fact, that B-type substrates (containing phosphorous-terminated steps) have shown defect formation at lower temperatures than A-type samples, seems to suggest the same picture.

Some aspects of the differences observed in the growth on substrates with A- and B-steps can also be understood from the generic (i.e.~not materials specific) properties of these steps. A-steps are group-III-terminated (In for InP) and B-steps are group-V-terminated (P for InP).  These steps have very different formation energies, which is reflected in their kinetics. The A-steps have a lower energy of formation than the B-steps. \cite{zhang96} and an In atom has a higher detachment barrier from a B-step than from an A-step. Hence, an In adatom can attach and detach from an A-step much more easily than from a B-step, so A-steps are smoother (straighter) than B-steps.\cite{bell99} This seems to be supported by the images presented earlier, and especially highlighted in Fig.~\ref{fig13} for surfaces misoriented by 0.4$^\circ$ along the A and B directions.

\subsection{Stable and unstable modes on misoriented surfaces}
\label{sec4B}

We can obtain a qualitative understanding of the morphological variations on A-surfaces of misoriented InP(001) by appealing to an earlier study of growth on  GaAs (001) by MOVPE.\cite{chua08}  The trend observed there on surfaces misoriented by 0$^\circ$ $(\pm0.02^\circ)$, 0.2$^\circ$, and 0.4$^\circ$ was islanding, step flow, incipient step bunching, and well-developed step bunching, respectively.  These results can be explained by a model that supposes that the decomposition of trimethylgallium is greater from above than below steps, which produces a net downhill current, as would an inverse Ehrlich--Schwoebel barrier.\cite{ehrlich66,schwoebel66}  The kinetics of the group-III adatoms were comparatively unimportant under these growth conditions.  However, the experiments reported here have been carried out under a much wider range of growth conditions, including large variations in the phosphine flow, which will enable a much more extensive study of the growth kinetics. For this reason, we will retain the full range of kinetic processes for both precursors and adatoms, including asymmetric step-edge incorporation.

\begin{figure}[t!]
\includegraphics[width=8cm]{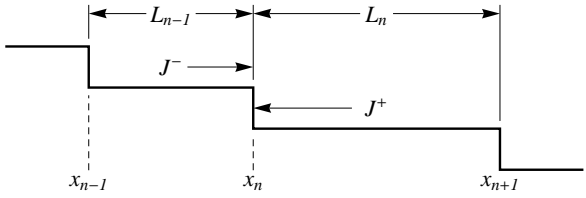}
\caption{\label{fig14}The positions $x_n$ of steps on a one-dimensional surface and the surface currents from above $(-)$ and below $(+)$ each step that drive step motion and determine the stability of the step train.}
\end{figure}

Our model\cite{chua08} is an extension of the basic BCF theory\cite{burton51} for the surface concentrations $n(x,t)$ of group-III precursors (i.e.~TMIn for InP) and $c(x,t)$ of group-III adatoms (i.e.~In) that includes the deposition rate of of precursors to the surface, their subsequent surface diffusion and possible desorption from the substrate, their decomposition at step edges, the release of group-III atoms from their precursors and the subsequent adatom surface diffusion.  The coupled reaction-diffusion equations for $n$ and $c$ that describe these processes are
\begin{align}
\label{eq1}
{\partial n\over\partial t}&=D_M{\partial^2n\over\partial x^2}-{n\over\tau}-\kappa n+F\, ,\\
\noalign{\vskip3pt}
{\partial c\over\partial t}&=D_A{\partial^2c\over\partial x^2}+\kappa n\, ,
\label{eq2}
\end{align}
where $D_M$ and $D_A$ are, respectively, the surface diffusion constants of the precursor and the adatoms, $\kappa$ and $\tau^{-1}$ are the decomposition and desorption rates, respectively of the precursor on the terrace, and $F$ is the effective deposition flux of the precursor.   The quantity $\kappa n$ therefore acts as an effective spatially-dependent deposition flux for group-III atoms. Although these equations are applicable for growth of a two-dimensional surface, in the interest of obtaining an analytic theory, we will confine our selves to a one-dimensional surface (Fig.~\ref{fig14}).  This, of course, neglects any step meandering which, as the images in the preceding section reveals, is an interesting issue in its own right.

The reaction-diffusion equations (\ref{eq1}) and (\ref{eq2}) are supplemented by boundary conditions at the leading and trailing edges that bound  each terrace. On the $n$th terrace, for which $x_n\le x\le x_{n+1}$, the boundary conditions for the precursors are
\begin{align}
\label{eq3}
D_Mn_x(x_n,t)&=\beta_M^+n(x_n,t)\, ,\\
\noalign{\vskip3pt}
-D_Mn_x(x_{n+1},t)&=\beta_M^-n(x_{n+1},t)\, ,
\label{eq4}
\end{align}
and for the adatoms,
\begin{align}
\label{eq5}
D_Ac_x(x_n,t)&=\beta_A^+[c(x_n,t)-c_0]\, ,\\
\noalign{\vskip3pt}
-D_Ac_x(x_{n+1},t)&=\beta_A^-[c(x_{n+1},t)-c_0]\, ,
\label{eq6}
\end{align}
where $c_0$ is the equilibrium adatom concentration at the step edge. The boundary conditions for the precursor stipulate that a molecule incident on step from above $(+)$ or below $(-)$ decomposes and the group-III atomic constituent incorporated at a rate proportional to $\beta_M^\pm$.  The boundary conditions for the group-III adatoms state that atoms incident on a step from above or below are incorporated into the solid at a rate proportional to $\beta_A^\pm$.

The morphological evolution of a misoriented surface is described by the changes in the positions $x_n$ of the steps as the result of absorbing material from surface currents (Fig.~\ref{fig14}),
\begin{equation}
{dx_n\over dt}=J^+(L_n)+J^-(L_{n-1})\, ,
\end{equation}
where $L_n=x_{n+1}-x_n$ and $L_{n-1}=x_n-x_{n-1}$ are the lengths of the upper and lower terraces, respectively, at $x_n$, and the surface currents $J^\pm$ are obtained by solving (\ref{eq1})--(\ref{eq6}) for each terrace in the system. The corresponding equations for the terrace lengths $L_n$ are
\begin{align}
{dL_n\over dt}&={dx_{n+1}\over dt}-{dx_n\over dt}\nonumber\\
\noalign{\vskip3pt}
&=J^+(L_{n+1})+J^-(L_n)-J^+(L_n)-J^-(L_{n-1})\, .
\label{eq8}
\end{align}
These equations have the stationary solution $L_n=L$ for all $n$, where $L=a/\tan\theta$ is the average terrace length determined by the misorientation angle $\theta$, where $a$ is the step height which, for InP(001) is the height of an In-P bilayer.  The central question in this paper is whether the regular train is stable to step bunching.  This can studied most expediently with a linear stability analysis by considering small deviations $\lambda_n$ from the regular step train:~$L_n=L+\lambda_n$.  By retaining terms only to first order in the $\lambda_n$ in the Taylor series of $J^\pm$,
\begin{equation}
J^\pm(L_{n\pm 1})=J^\pm(L)+{dJ^\pm\over dL}\bigg|_0\lambda_{n\pm1}\, ,
\end{equation}
where the subscript on the derivatives indicates evaluation with the regular step train, the linearized form of (\ref{eq8}) is obtained as
\begin{equation}
{d\lambda_n\over dt}={dJ^+\over dL}\bigg|_0(\lambda_{n+1}-\lambda_n)+{dJ^-\over dL}\bigg|_0(\lambda_n-\lambda_{n-1})\, .
\end{equation}
With the step displacements expressed as Fourier modes,
\begin{equation}
\lambda_n=u_k e^{inkL-i\omega_k t}\, ,
\end{equation} 
the growth or decay rate $R_k$ of the $k$th mode, which is the real part of $\omega_k$, is
\begin{equation}
R_k=2\sin^2\bigl({\textstyle{1\over2}}kL\bigr){dJ_s\over dL}\bigg|_0\, ,
\label{eq12}
\end{equation}
where $J_s(L)=J^-(L)-J^+(L)$. The stability of the regular step train is determined by the sign of $R_k$ which, in turn, is determined by the sign of $J_s^\prime(L)$:~if $R_k<0$, the regular step train is stable, while if $R_k>0$, it is unstable to step bunching.  The most unstable mode corresponds to $k=\pi/L$, with eigenvector $(1,-1,1,-1,\ldots)$, so step bunching is initiated by the pairing of adjacent steps.

\subsection{Stability of step flow on misoriented InP(001)}
\label{sec4C}

The decay rate $R_k$ is determined from the stationary solution of (\ref{eq1})--(\ref{eq6}).  The general solutions to (\ref{eq1}) and (\ref{eq2}) are straightforward to determine, but the subsequent calculation of the currents and their derivatives are quite lengthy.  Therefore, the generation and manipulation of these solutions have been carried out in {\sc Mathematica}.\cite{wolfram}  The rates $\Gamma_n$ of all kinetic processes in our model have an Arrhenius form, $\Gamma_n=\nu_n e^{-E_n/k_BT}$, so determining $R_k$ necessitates assigning values to the prefactor $\nu_n$ and barrier $E_n$ for each process.  As our model does not explicitly include the group-V kinetics, these parameters have an implicit dependence on the phosphine flow rate.

\begin{figure*}[t!]
\includegraphics[width=7cm]{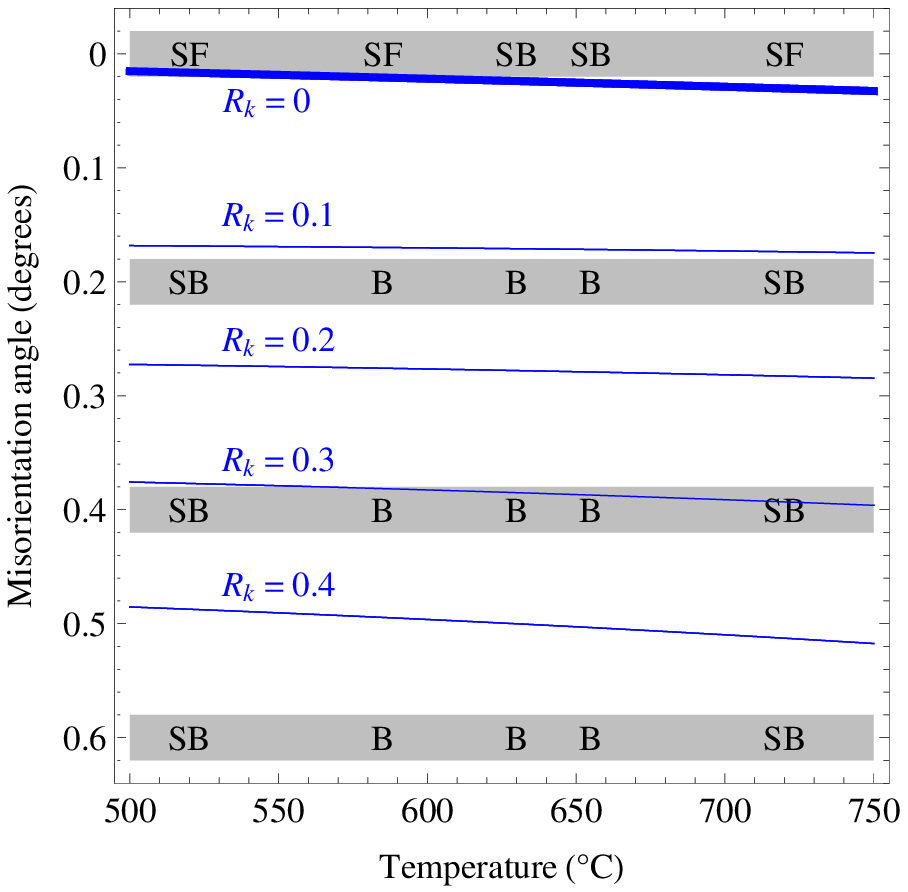}\hskip1cm\includegraphics[width=7cm]{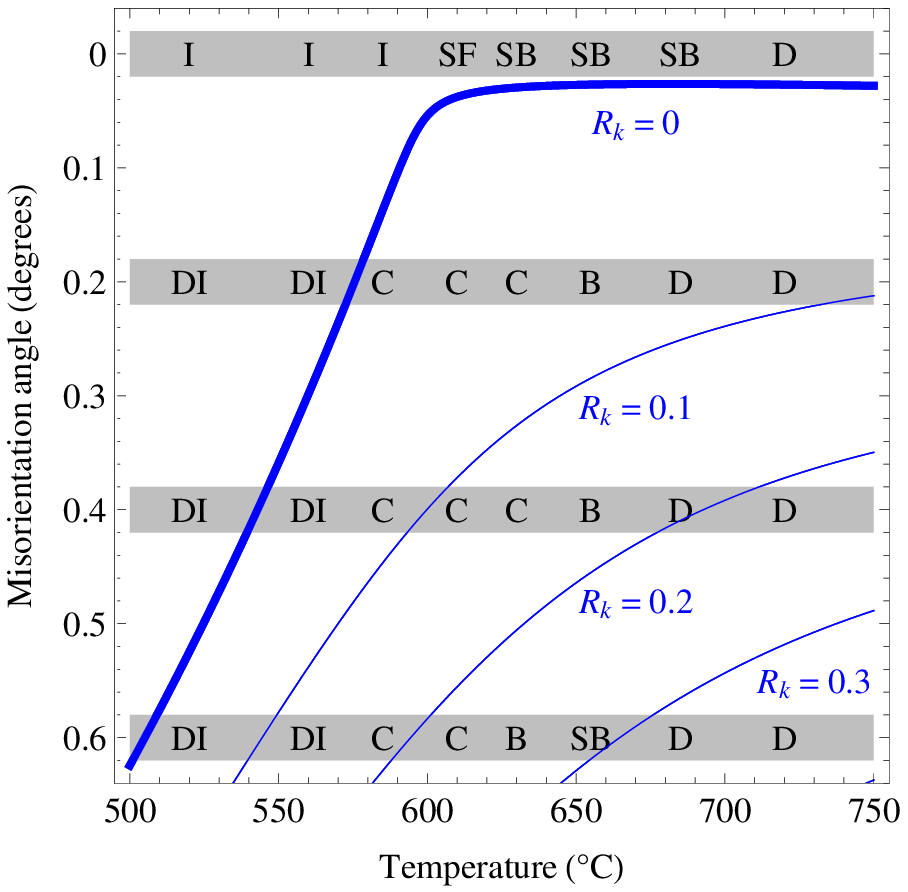}
\caption{\label{fig15}(Color online) The morphologies in Table~\ref{table1} for different misorientation angles (toward [111]A) for $R=450$ (left panel) and $R=150$ (right panel).  The contours of the decay rate in (\ref{eq12}), which are indicted in blue, have been calculated with the Arrhenius parametrizations in Table~\ref{table5}. The emboldened curves correspond to $R_k=0$, which separates regions of stable $(R_k<0$ and $(R_k>0$) step flow, and the gray regions indicate the accuracy of $\pm0.02^\circ$ on the misorientation angles.}
\end{figure*}

\begin{table}[b!]
\renewcommand{\arraystretch}{1.2}
\caption{\label{table5} Frequency prefactors $(\nu)$ and energy barriers $(E)$ for the Arrhenius rates of the kinetic parameters in (\ref{eq1})--(\ref{eq6}) used for the fits in Fig.~\ref{fig15}.  The parameters whose values are enclosed in parentheses do not affect the results in Fig.~\ref{fig15}(a), even for large variations.}
\medskip
\begin{tabular}{|l|c|c|c|c|}
\hline
& \multicolumn{2}{c}{$R=150$} & \multicolumn{2}{|c|}{$R=450$} \\
\cline{2-5}
& $\nu$ (s$^{-1}$) & $E$ (eV) & $\nu$ (s$^{-1}$) & $E$ (eV) \\
\hline
$D_M$ & $10^{14} $ & 0.1 & $10^{14}$ & 0.2 \\
$D_A$ & $10^{12} $ & 0.44 & $10^{12}$ & (0.6) \\
$\kappa$ & $10^{10} $ & 0.8 & $10^{10}$ & 0.8 \\
$\tau^{-1}$ & $10^{10} $ & 1.2 & $10^{10}$ & 1.2 \\
$\beta_M^-$ & $10^{12} $ & 0.8 & $10^{12}$ & 0.8 \\
$\beta_M^+$ & $10^{12} $ & 1.0 & $10^{12}$ & 1.0 \\
$\beta_A^-$ & $10^{12} $ & 0.75 & $10^{12}$ & (1.0) \\
$\beta_A^+$ & $10^{12} $ & 0.5 & $10^{12}$ & (1.0) \\
\hline
\end{tabular}
\end{table}

Our model will be used to analyze the morphologies in Table~\ref{table1}, as this provides the most comprehensive temperature-misorientation (toward [111]A) matrix of data.  These data and our fits are shown in Fig.~\ref{fig15}, with the Arrhenius parameters for the kinetic coefficients appearing in (\ref{eq1})--(\ref{eq6}) compiled in Table~\ref{table5}.  The focus of most of our discussion will be on processes occurring at step edges, but the prefactor for the surface diffusion of TMI merits some comment.  The value $\nu=10^{14}$, which may seem quite high in comparison with the range $10^{12}$--$10^{13}$ typically used in such theories and simulations, actually conforms with estimates\cite{schunack02} for large molecules.  Such large prefactors are due to jumps over several lattice cites, which is in contrast to the usual assumption of jumps of a single lattice unit. Otherwise, the precise values in Table~\ref{table5} are not as important as the trends they reveal as the growth conditions are varied.

We first consider the morphologies at high phosphine flow rate $(R=450)$.  The trends in the left panel of Fig.~\ref{fig15} are similar to those seen on GaAs(001) (cf.~Fig.~3 of Ref.~[\onlinecite{chua08}]), which was also grown under group-V-rich conditions.  The direct incorporation of the adatom from the precursor occurs at a greater rate from above the step than below for both systems, but there is no asymmetry in the adatom step incorporation kinetics.  Indeed, large (i.e.~several tenths of electron volts) variations of the energy barriers for the adatom diffusion and incorporation only weakly affect on the stability calculation, provided that the incorporation kinetics at the step edge are symmetric. We have enclosed the barriers for these processes in parentheses in Table~\ref{table5} to indicate this fact.  For these growth conditions, the instability of step flow to step bunching in InP(001) and GaAs(001) is driven by an effective negative step-edge barrier induced by the asymmetric precursor incorporation kinetics.\cite{note}

An altogether different scenario emerges from the stability analysis at low phosphine flow rates, as can be seen immediately in the right panel of Fig.~\ref{fig15}.  At lower temperatures ($T_G=520^\circ$C and 560$^\circ$C) the morphology of all substrates shows either discrete or diffuse islands (Table~\ref{table1}).  Discrete islands indicate a growth regime far removed from step flow. The adatom density on the terrace is large, leading to a high probability of island nucleation and growth on the terraces.  As the temperature is increased, the adatom density on the terraces decreases, so the growth nucleation and growth of islands decreases accordingly, with the islands showing some coalescence with steps, thereby indicating a transition regime between discrete islanding and step flow. The separation of stable and unstable regions of step flow is expected to be qualitatively different from that obtained with higher phosphine flow rates.   Table~\ref{fig5} shows that, while the asymmetry in the molecular step-edge incorporation rates has been maintained, an asymmetry in the {\it atomic} step-edge incorporation rates has been introduced, but in {\it opposition} to the molecular incorporation rates.  This is the reason for the qualitative difference in shape of the curve that separates the stable and unstable regions $(R_k=0)$.

Our calculations show that the profile of the curve $R_k=0$ separating the stable and unstable regions is very sensitive to the atomic parameters. 
With decreasing temperature, the net incorporation current at steps becomes more negative, while that due to precursor incorporation becomes more positive.  For certain growth conditions, the atomic current dominates, which suggests the onset of another instability through the appearance of mounds, a fact that has also been noted by Vladimirova {\it et al.}\cite{vladimirova00} In fact, mounds have been observed on InP(001) during metalorganic molecular-beam\cite{cotta93} and chemical-beam\cite{bortoleto02} epitaxy.  The interesting point from our perspective is that this morphology is triggered by the low phosphine flow rate and can therefore be avoided.

\section{Conclusions}
\label{sec5}

Homoepitaxy on perfectly oriented substrates resulted in the most homogenous surfaces -- stable step flow was observed in the majority of cases over some temperature range, with step-bunching composed of a maximum of three steps observed near 650$^\circ$C. The modification of the V/III ratio had the most striking effect on the surface morphology. For $R=150$, a growth temperature below 600$^\circ$C was found to be insufficient to create useful surface organization for any misorientation of the substrate (the exception being the substrate with A-steps with a misorientation of 0.6$^\circ$, where regular step organization was observed at 585$^\circ$C). Temperatures above 685$^\circ$C led to defects related to three-dimensional growth. The mid-range growth temperatures provided more of a distinction between various substrates. Step-bunching was observed in most of the cases but, depending on the surface misorientation and exact growth temperature, different features were formed on the surface. When examined in detail, the overlap of crystallographic steps could take diverse forms, like clustering of just two monolayers (on perfectly oriented substrates), well organized, periodic micron-scale bunching (cliffs observed on A-type substrates at 655$^\circ$C), or a braid-like bunching of up to 40 monolayers (on all vicinal surfaces grown at 630$^\circ$C). This limited the optimal growth temperature to a quite narrow window of about 100$^\circ$C. By changing the phosphine flow to $R=450$, most of the samples were eliminated by that restriction -- stable, organized step-flow was observed even at 520$^\circ$C and good surface morphology was obtained for growths conducted at temperatures of 720$^\circ$C.

There are several aspects of the surface morphology that can be addressed with modelling. The basis of our theoretical work\cite{chua08} has been a continuum model of the Burton, Cabrera, and Frank\cite{burton51} type.  This allows an analytic solution to be obtained and a linear stability analysis to be carried out, but cannot address the variety of morphologies seen for fully developed step bunching, for which the solution of the two-dimensional problem is required.  There are two approaches.  One is to use a continuum formulation, for example, one based on the phase-field method.\cite{yu11} This has the advantage of building the extended terrace length scales on the substrate from the outset, but suffers from the difficulty of including the details of step kinetics while retaining a direct connection to the atomistic parameters.  Kinetic Monte Carlo (KMC) simulations provide an obvious alternative in that the atomistic processes can be included to whatever level of detail required, but the large terrace lengths present a computational challenge for traditional formulations.  One alternative is to use a hybrid scheme that combines the flexibility of the KMC approach and the extended length and time scales of continuum methods.\cite{schulze04} These methods are currently under investigation.

\section*{Acknowledgements}

This research was supported by the Irish Higher Education Authority Program for Research in Third Level Institutions (2007-2011) via the INSPIRE programme, by the Science Foundation Ireland under grants 05/IN.1/I25, 10/IN.1/I3000 and 07/SRC/I117 and by the European 7th Framework Programme under grant 228033 (MODE-GAP). The authors are grateful to Dr.~K.~Thomas for support of the MOVPE system. D.D.V. thanks B. A. Joyce for many helpful discussions about precursor chemistry.


\begin{thebibliography}{99}
\bibitem{note} It should be noted that the detailed evolution at high $T_G$, high V/III and small misorientation (Fig.~\ref{fig15} left) is actually not fully taken into consideration in our model, and will probably need a more comprehensive 2D model of surface dynamics, including e.g. phosphine desorption.
\end{thebibliography}
\end{document}